\documentclass[%
 reprint,
nofootinbib,
 amsmath,amssymb,
 aps,
]{revtex4-2}

\usepackage{graphicx}% Include figure files
\usepackage{dcolumn}% Align table columns on decimal point
\usepackage{xcolor}
\usepackage{color}
\usepackage{bm}% bold math
\usepackage{orcidlink}
\begin{document}

\preprint{APS/123-QED}

\title{Reconstructing the LISA massive black hole binary population via iterative kernel density estimation}% Force line breaks with \\
%\thanks{corresponding author: jsadiq@sissa.it}%
%\author{{Jam Sadiq \textsuperscript{1, 2} \thanks{Corresponding author, \email:jsadiq@sissa.it}} \and
%  Kallol Dey\textsuperscript{4} \and
%  Thomas Dent\textsuperscript{3} \and
%  Enrico Barausse\textsuperscript{1,2}}

%\affiliation{\textsuperscript{1}SISSA, Via Bonomea 265, 34136 Trieste, Italy and INFN Sezione di Trieste}
%\affiliation{\textsuperscript{2}IFPU - Institute for Fundamental Physics of the Universe, Via Beirut 2, 34014 Trieste, Italy}
%\affiliation{\textsuperscript{3}IGFAE, University of Santiago de Compostela, E-15782 Spain}
%\affiliation{\textsuperscript{4}School of Physics, Indian Institute of Science Education and Research Thiruvananthapuram, Maruthamala PO, Vithura, Thiruvananthapuram 695551, Kerala,   India}

\author{Jam Sadiq\orcidlink{0000-0001-5931-3624}\textsuperscript{1, 2}}\thanks{Corresponding author, email: jsadiq@sissa.it}
\author{Kallol Dey\orcidlink{0000-0003-2145-1145}\textsuperscript{3}}
\author{Thomas Dent\orcidlink{0000-0003-1354-7809}\textsuperscript{4}}
\author{Enrico Barausse\orcidlink{0000-0001-6499-6263}\textsuperscript{1, 2}}

\affiliation{\textsuperscript{1} SISSA, Via Bonomea 265, 34136 Trieste, Italy and INFN Sezione di Trieste}
\affiliation{\textsuperscript{2} IFPU - Institute for Fundamental Physics of the Universe, Via Beirut 2, 34014 Trieste, Italy}
\affiliation{\textsuperscript{3} School of Physics, Indian Institute of Science Education and Research Thiruvananthapuram, Maruthamala PO, Vithura, Thiruvananthapuram 695551, Kerala, India}
\affiliation{\textsuperscript{4} IGFAE, University of Santiago de Compostela, E-15782 Spain}

%\author{Enrico Barausse \textsuperscript{1, 2}}
 %\affiliation{IFPU - Institute for Fundamental Physics of the Universe, Via Beirut 2, 34014 Trieste, Italy}
%\affiliation{SISSA, Via Bonomea 265, 34136 Trieste, Italy and INFN Sezione di Trieste}

% This line break forced with \textbackslash%\textbackslash
%}%

%\affiliation{
%\affiliation{
%\affiliation{%
% This line break forced with \textbackslash%\textbackslash
%}%

%\collaboration{CLEO %Collaboration}%\noaffiliation

\date{\today}% It is always \today, today,

\begin{abstract}

Reconstructing the properties of the astrophysical population of binary compact objects in the universe is a key science goal of gravitational wave detectors. This goal is hindered by the finite strain, frequency sensitivity and observing time of current and future detectors. This implies that we can in general observe only a selected subset of the underlying population, with limited event statistics, and also nontrivial observational uncertainties in the parameters of each event. In this work, we will focus on observations of massive black hole binaries with the Laser Interferometer Space Antenna (LISA). If these black holes grow from population III star remnants (``light seeds''),  a significant fraction of the binary population at low masses and high redshift will be beyond LISA's observational reach; thus, selection effects have to be accounted for when reconstructing the underlying population. 
Here we propose an iterative, kernel density estimation (KDE)-based non-parametric method, in order to tackle these statistical challenges in reconstructing the astrophysical population distribution from a finite number of observed signals over total mass and redshift.
We test the method against a set of simulated LISA observations in a light seed formation scenario. We find that our approach is successful at reconstructing
the underlying astrophysical distribution in mass and redshift, except in parameter regions where zero or order(1) signals are observed.  
\end{abstract}

\maketitle

\section{\label{sec:level1} Introduction}

The birth of gravitational wave (GW) astronomy with the first detection of the binary black hole merger GW150914~\cite{LIGOScientific:2016aoc} and the subsequent more than a hundred detections~\cite{LIGOScientific:2018mvr,LIGOScientific:2020ibl,ligo_scientific_collaboration_and_virgo_2021_5117970,KAGRA:2021vkt} by the LIGO-Virgo-KAGRA (LVK) collaboration
have opened a novel perspective on the cosmos, providing not only evidence that compact objects coalesce, but also a way to extract the statistical properties of their astrophysical population~\cite{KAGRA:2021duu}.

Because GW interferometers have intrinsically limited sensitivity in strain and frequency, they typically access only a fraction of the 
compact object binaries in our past light cone. On the ground, these selection effects are the main reason of the uncertainties in the reconstruction of the mass function of stellar origin black hole binaries~\cite{Mandel:2018mve,Gaebel_2019,Vitale_2021}. Previous methods, such as those outlined in~\cite{Finn:1992xs,Finn:1995ah}, typically factor out the dependence of the extrinsic angles on the event signal-to-noise ratio (SNR). Events can be subsequently thresholded based on the SNR of an optimal source with the same intrinsic parameters. Recent state-of-the-art GW population studies include selection biases in some approximate form~\cite{Taylor:2018iat,LIGOScientific:2018jsj,Roulet:2018jbe}, and have been further improved with different approaches~\cite{Tiwari:2017ndi,sensitivityDan,Lorenzo-Medina:2024opt}. Sensitivity estimates based on SNR thresholding have been applied in \cite{LIGOScientific:2018jsj,KAGRA:2021duu}; a recent comparison with search injection (simulated signal) campaigns in real LIGO-Virgo noise is presented in \cite{Essick:2023toz}.

While these uncertainties will be mitigated with updates to the current facilities
and with next-generation ground based detectors such as the Einstein Telescope~\cite{ET0106C2011, Maggiore:2019uih,Branchesi:2023mws} or Cosmic Explorer~\cite{2021arXiv210909882E,Hall:2022dik}, it is expected that even the latter may miss a significant fraction of the black hole population at high redshift (see e.g.\ Fig.~3 of \cite{Pieroni:2022bbh}). Moreover, even when high-redshift binaries are observed, their distance determination will be significantly degraded, due to statistical errors (caused by the low signal-to-noise ratios) and weak-lensing systematics, which are the dominant source of error on the distance at high $z$~\cite{2010PhRvD..81l4046H,Tamanini:2016zlh}. 
 Another potential source of systematics is given by environmental effects. However, those are expected to be negligible for massive black hole binaries in the LISA band~\cite{Barausse2014, Barausse:2014pra}, which are the focus of this paper. This makes the reconstruction of the distribution of the underlying population in mass and redshift even more challenging.

It should also be noted that the black hole population of the universe does not only consist of remnants of stellar evolution, such as those observed by the LVK collaboration. Massive black holes (MBHs) with masses 
$10^5$--$10^9\,M_\odot$ have been observed
at the center of most large elliptical galaxies~\cite{Gehren84,Kormendy1995}, and also
in some lower mass systems~\cite{reines11,reines13,Baldassare2019}.
These MBHs, when they accrete, are also the engine of  Active Galactic Nuclei (AGNs) and quasars~\cite{2002apa..book.....F}, and as such they constitute
a crucial ingredient of galaxy formation and evolution, since they
are believed to exert feedback (via radiation, disk winds or jets)
on their galactic hosts~\cite{Croton2006,2008ApJS..175..390H,Bower2006}. Because galaxies form hierarchically, through the merger of smaller systems into bigger ones, the MBHs at the center of galaxies are also expected to merge, emitting GW signals of frequencies lower than those accessible from the ground (where interferometers are limited by seismic noise).

The origin of MBHs is still to some extent a mystery. Several models 
for the black hole ``seeds'' of the MBH population at high redshift $z\gtrsim 10-15$
have been proposed, but can be broadly divided into two classes, ``heavy seeds'' of masses $\sim 10^5\,M_\odot$
or ``light seeds'' of masses $10^{2}-10^3\,M_\odot$ (see e.g.~\cite{Latif:2016qau} for a review). The latter may even survive to low redshift to provide a population
of intermediate mass black holes between MBHs and stellar-origin black holes~\cite{IMBH}. A mixing~\cite{Sesana:2010wy,Toubiana:2020lzd} of these light and heavy populations is also possible, if not probable. The growth of the MBH seeds, via accretion and mergers, is also subject to considerable uncertainties.
On the theory side, MBHs are tiny compared to galactic scales, and their sphere of influence is difficult to resolve in simulations. Also, many processes that are crucial for their evolution, such as 
star formation, supernova feedback, AGN feedback and accretion itself,
are not yet fully understood from first principles. For this reason,
MBH evolution, especially at high redshift, is heavily affected by 
the ``sub-grid'' prescriptions used to describe these processes in hydrodynamic simulations (cf.\ e.g.~\cite{DiMatteo2005, Dubois_2012, Vogelsberger:2014dza, Schaye2015, Volonteri_2021, Tremmel2017, Pontzen2017, Nelson2019, Ricarte:2018mzn}),
or is followed by using semi-analytic galaxy evolution models~\cite{Volonteri:2002vz, Somerville:2008bx, Barausse:2012fy, Sesana:2014bea, Bonetti:2018tpf, Tremmel2015, Tremmel2018, Tremmel2018b, Barausse:2020mdt, Barausse:2023yrx}.
On the observational side, with electromagnetic telescopes, MBHs are hard to observe at high redshift, when their growth and AGN activity are expected to be strongest.

These difficulties open a significant potential discovery window for space-borne detectors of GWs and for pulsar-timing arrays. Regarding the latter, there is now evidence of a common correlated signal in the
data of several experiments~\cite{EPTA:2023fyk, Tarafdar:2022toa, NANOGrav:2023gor, Reardon:2023gzh, Xu:2023wog}, which could be the long-expected stochastic background of GWs in the nHz band from a population of MBHs of masses $\gtrsim 10^8\,M_\odot$~\cite{EPTA:2023xxk, NANOGrav:2023hfp}. At higher frequencies, the Laser Interferometer Space Antenna (LISA)~\cite{LISA2013,LISA2017} is scheduled to be launched in 2034 and will observe MBH binaries of masses $10^4-10^7\,M_\odot$ in the band from $\sim 0.01$\,mHz to $\sim 0.1$\,Hz. The exact redshift reach of LISA depends on the binary mass, with masses $10^5-10^6 M_\odot$ observable up to $z\sim 20$ or more. However,
binaries with masses $\lesssim 10^4\,M_\odot$ are harder to observe at high redshift, and the inference of their distance is more challenging (due to the lower signal-to-noise ratio and to weak lensing~\cite{2010PhRvD..81l4046H, Tamanini:2016zlh}). This is exactly the mass range of light seeds/intermediate mass black holes, whose existence is still elusive~\cite{IMBH}. 

While we expect LISA to offer unique insights into this black hole population, if it exists, the reconstruction of its  distribution in mass and redshift will therefore be subject to various statistical challenges. First, the finite number of observed events; second, the presence of significant selection effects, because many such systems will be undetected (both from space and from the ground-based detector network); 
third, the significant uncertainties in the parameters of any given merger event due to detector noise, particularly affecting distance and redshift measurements.

In this paper, we propose a method to deal with these statistical issues and correct or reduce associated biases in the reconstructed population. In more detail, we will use the results of a semi-analytic galaxy formation model in which MBHs evolve from light seeds~\cite{Barausse:2023yrx}, forming as the remnants of the explosion of Population III stars at high $z$~\cite{Madau2001,Madau2014}, to simulate a population of MBH binary mergers. For each merger in our past light cone, we will assess its detectability with LISA, and for each detectable system we will obtain posterior samples for all source parameters. These samples include the effect of weak lensing on the distance estimation~\cite{2010PhRvD..81l4046H,Tamanini:2016zlh}. Based on these simulated samples, we will then attempt to reproduce the distribution of the astrophysical population in a two-dimensional mass-redshift parameter space, using an adaptive kernel density estimation (KDE) method and correcting for selection effects due to the finite LISA sensitivity.
We will show that our method provides an unbiased reconstruction of the MBH mass and redshift distribution function, allowing for shedding light on the astrophysical formation channels of MBH binaries and on the prevalence of intermediate mass black holes in the universe.

This paper is organized as follows: in Sec.~\ref{sec:dataanalysis-section}, we describe the generation of mock (simulated) data and our analysis methods for this work; in Sec.~\ref{sec:result-section}, we report and discuss our results, and in Sec.~\ref{sec:conclusion-section} we describe conclusion and future work.

\section{Simulated data}
\label{sec:dataanalysis-section}
\subsection{Astrophysical population model and detectability}
\label{ss:catalogdata}

To describe the astrophysical population of MBH binaries, we consider 
the light-seed semi-analytic model presented in Ref.~\cite{Klein:2015hvg} (and based in turn on Refs.~\cite{Barausse:2012fy,Sesana:2014bea,Antonini_Barausse2015,Antonini2015}, to which we refer for more details on the model's physical content). This model was referred to as ``popIII-d (K+16)'' in Ref.~\cite{Barausse:2023yrx}, where it was shown to agree with the amplitude estimate for the stochastic background of GWs in the band of pulsar-timing arrays~\cite{EPTA:2023fyk, Tarafdar:2022toa, NANOGrav:2023gor, Reardon:2023gzh, Xu:2023wog}. Note that the ``popIII-d (K+16)'' of Ref.~\cite{Barausse:2023yrx}
does include delays between galaxy and MBH mergers. We stress, however, that our goal here is not to reconstruct the distribution of the binaries at formation, but the distribution at merger. For this task, the delays need only to be included in the simulated data used for the KDE reconstruction. 
\begin{figure}[tbp]
    \includegraphics[width=0.48\textwidth]{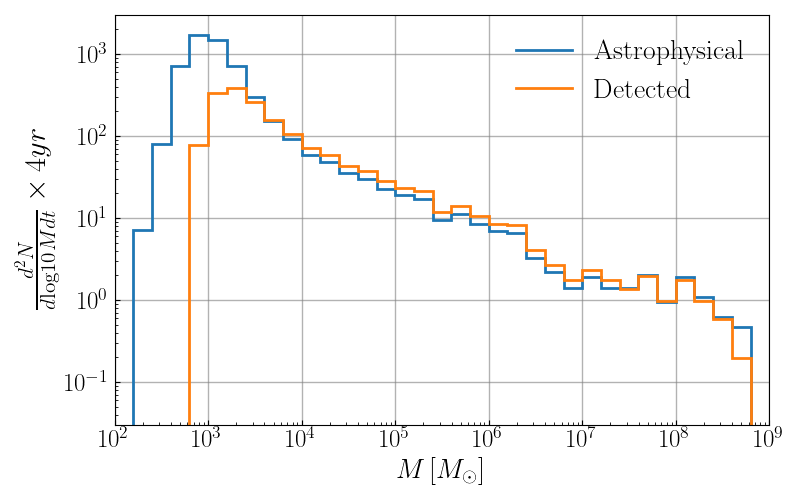}
     \includegraphics[width=0.48\textwidth]{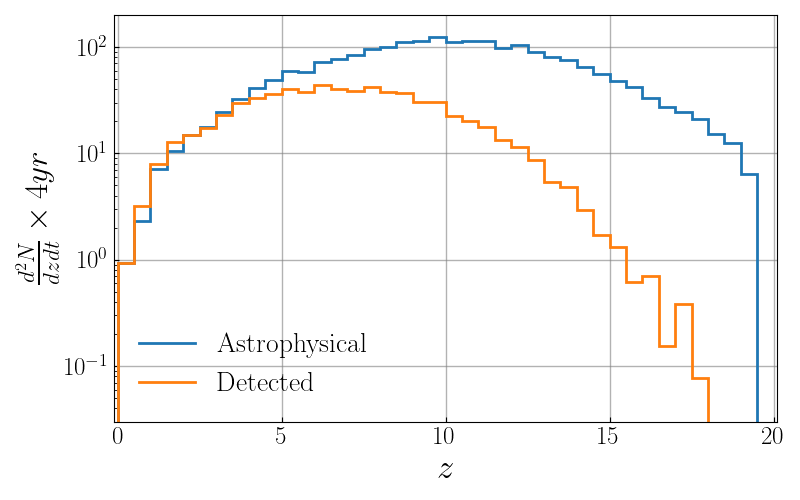}    
     \caption{The differential rate of all mergers (light blue line) and of detected mergers (orange line), for the light-seed ``popIII-d (K+16)'' model that we consider in this work. Top: density over source frame total mass.  Bottom: density over redshift.}
    \label{fig:catalogdistribution}
\end{figure}

From the astrophysical population, we obtain simulated catalogs of merger events in the band of the LISA mission, for which we assume a duration of 4 years. To simulate the GW signal, we follow  Ref.~\cite{Barausse:2023yrx} and use the non-precessing IMRPhenomHM model~\cite{London:2017bcn}, which describes GW signals from binaries with aligned spins and also includes contributions from higher modes. 
In more detail, our simulated GW signals are described by the following 11 parameters: the masses ($M_1$ and $M_2$), the spins components along the orbital axis ($\chi_1$ and $\chi_2$), the time of coalescence ($t_c$), the luminosity distance ($D_L$), the inclination angle ($\iota$), the sky location described by the longitude ($\lambda$) and latitude ($\zeta$), the polarization angle ($\psi$) and the phase at coalescence ($\phi$). The masses, spins and the distance are obtained from the astrophysical model, the merger time is drawn uniformly between 0 and 4 yr (the mission duration), while the angles are drawn from an isotropic distribution.

For the instrumental response, we use the full LISA response function, following~\cite{Marsat:2018oam}.
The Time Delay Interferometry (TDI)~\cite{Estabrook:2000ef, Armstrong_1999, Dhurandhar:2001tct, Tinto:2004wu} observables A, E, and T are calculated for each merger event. As the T channel has very low signal content, it can be excluded, and we use the A and E channels for all further analyses. Once the signal of an event has been simulated, we estimate the optimal SNR $\rho$ to assess detectability, using 
\begin{equation}
    \rho^2 = (h|h) \,,
\end{equation}
where $h(t)$ is the GW signal, and the inner product $\left(\cdot|\cdot\right)$ is defined as
\begin{equation}
    (a|b) = 4\, \text{Re} \int_0^\infty \frac{\Tilde{a}^*(f)\Tilde{b}(f)}{S_n(f)}\,
    df \,,
\end{equation}
where $S_n(f)$ is the noise power spectral density. We employ the SciRDv1 noise model \cite{LISA_SciRDv1}, augmented with an unresolved white dwarf background from Galactic binaries \cite{Babak2017, Robson:2018ifk}. Our computation of $\rho$ accounts for the significant cross-terms between different emission modes (multipoles)~\cite{Pitte:2023ltw}. For the events with $\rho>8$, we perform detailed parameter estimation (PE) as described in the next section. 

Figure \ref{fig:catalogdistribution} presents the expected number of mergers predicted by the model over 4 years, plotted as distributions over source mass (top) and redshift (bottom). 
The light blue line represents the total number of mergers (\emph{astrophysical} distribution), while the orange line indicates the mergers
detectable by LISA (\textit{detected} distribution), assuming a 4-year mission and SNR threshold of 8. Because of the way
the semi-analytic model's results are post-processed to produce simulated catalogs of black hole mergers, the latter occur at discrete redshift values. To avoid unrealistic distribution artifacts due to this, when plotting the $z$ distribution we apply a small Gaussian scatter (negligible compared to measurement errors).
The expected total number of BBH mergers taking place in 4 years is 1425, as compared to 338 detected events. As can be seen, LISA is expected to miss a significant fraction of the merger population, making it a nontrivial task to reconstruct the astrophysical population.  By contrast, this is typically not the case for heavy-seed population models, where virtually the entire merger population is detectable by LISA~\cite{Klein:2015hvg,Barausse:2023yrx}. The {\it actual} number of mergers taking place, or detected, in the LISA observation time is a stochastic variable: in this work we will simulate one realization of 4 years of LISA data, and attempt to reconstruct underlying properties of the astrophysical model from these observations.

\begin{figure}[btp]
    \includegraphics[width=0.48\textwidth]{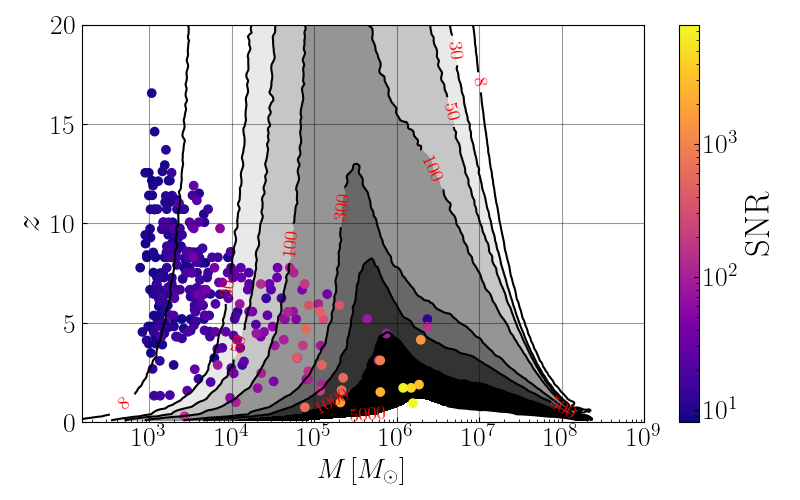}
    \caption{True values of source-frame mass and redshift for one realization of the mergers occurring in 4 years within the light-seed model ``popIII-d (K+16)'' employed in this work. The contours show optimal LISA SNR values assuming equal masses, no spins, 4 years of observations and averaging over the extrinsic angle parameters, while the color code denotes the optimal SNR computed with the actual source parameters.}
    \label{fig:true-SNRdetectedLensedevent4years}
\end{figure}
 
As another way to visualize the astrophysical population and its detectability, we show in Fig.~\ref{fig:true-SNRdetectedLensedevent4years} 
the true parameters of one realization of the mergers for 4 years of observation, within the ``popIII-d (K+16)'' light-seed model considered in this work. Also shown are contours of constant sky-, inclination- and polarization-averaged optimal SNR in the parameter plane of source-frame total mass $M$ and redshift $z$, assuming a 4-year observation period, zero component spins and a mass ratio $q=1$. For each source, we also show by the color code the actual optimal SNR, computed with the mass ratio and spins predicted by the astrophysical model, with the sky position, inclination and polarization angle and observation time values of the catalog realization, and with the luminosity distance predicted by the astrophysical model with the addition of a weak-lensing contribution as detailed in the next section. With this realization of a detected event set, we can hope to reconstruct the astrophysical distribution only between total masses of $\sim$$10^3 - \mathrm{few} \times 10^6\,M_\odot$ and redshifts $\sim 0-15$.

\subsection{Parameter Estimation}
\label{ss:pe_method}
 
We use the GW signals described in the previous section as injections on which we perform Bayesian PE, following  Ref.~\cite{Barausse:2023yrx}. Although the detected GW signals from MBH binaries are expected to overlap in the LISA band, we consider each signal in isolation to simplify the analysis. We also neglect the noise contribution to the injection (noiseless limit) and ignore possible data artifacts such as gaps and glitches~\cite{Dey:2021dem,Baghi:2019eqo,Spadaro:2023muy}. The posteriors are expected to only undergo a shift up to the statistical error with the inclusion of Gaussian noise in the injection, while the variance measurements are expected to be largely unaffected~\cite{Rodriguez:2013oaa,Sampson:2013lpa}.

We also consider the effect of weak lensing, which is expected to have an impact on the measurement of $D_L$ by introducing a random change in the GW signal amplitude.  The weak-lensing error is estimated as~\cite{Tamanini:2016zlh}
\begin{equation}
  \label{eq:sigma_wl}
    \sigma_{\rm wl}(z) = D_L \times 0.066 \left(\frac{1-\left(1+z\right)^{-0.25}}{0.25} \right)^{1.8},
\end{equation}
assuming a Planck 2015 $\Lambda$CDM cosmology. Considering $\sigma_{\rm wl}^2$ as the variance, we draw $\delta D_L$ from a normal distribution $\mathcal{N}(0,\sigma_{\rm wl}^2)$ and add this term to the $D_L$ predicted by our astrophysical model, so that the simulated LISA signal includes the effect of weak lensing. 

We infer the posterior parameter distribution for every source with optimal SNR larger than 8, using the \textit{LISAbeta} code~\cite{marsat2024lisabeta, Marsat:2020rtl}. The sampled parameters are total mass ($M$), mass ratio ($q$), $\chi_1$, $\chi_2$, $t_c$, $D_L$, $\iota$, $\lambda$, $\zeta$, $\psi$ and $\phi$.  The priors for the parameters $M$, $q$, $\chi_1$, $\chi_2$, $t_c$, and $D_L$ are chosen to be uniform. For $\iota$ and the sky location, we consider uniform priors on $\cos\iota$, $\lambda$ and $\sin \zeta$. For the sampling, we use \textit{ptemcee}~\cite{Vousden_2015}, an implementation of the Parallel Tempering Markov Chain Monte Carlo algorithm~\cite{B509983H}. In order to reduce computation times, we initiate the sampler at the ground truth of the sources, with an initial covariance matrix for proposal distributions coming from the Fisher matrix. This allows us to directly sample regions with high posterior probabilities while reducing the computation cost. Therefore, the Fisher matrix is never used for approximating the posteriors, which are instead fully Bayesian. The proposal distributions are also tuned to include secondary peaks in the sky locations~\cite{Mangiagli:2022niy}, which may have been missed due to our Fisher-based initialization.

The sampler is configured to run with 64 walkers and 10 ``temperatures'', and generates 8000 samples per source. The frequency bounds for the likelihood integral are taken as $f_{\rm min} = 10^{-4}$~Hz and $f_{\rm max}=0.5$~Hz. 
 Typical resulting fractional uncertainties in the redshifted mass $M_z \equiv M(1+z)$ and in the redshift $z$ are
$\sim 10^{-3}$ and $\sim 0.25$, respectively~\cite{Barausse:2023yrx}. 
Thus, as $M_z$ is precisely determined, errors in the \emph{source} total mass $M$ are strongly correlated with errors in $D_L$ or $z$.  These large and correlated event-level uncertainties are a challenge for accurate population reconstruction.

\subsection{Population reconstruction via iterative KDE}
\label{ss:iterkde}

Ref.~\cite{Sadiq:2021fin} demonstrated the use of adaptive Kernel Density Estimation (KDE)~\cite{awkdecodelink} to study rates and populations of binary black holes observed with the current GW detector network~\cite{KAGRA:2021vkt}. In~\cite{Sadiq:2023zee} we further developed the adaptive KDE method to account for measurement uncertainties in a self-consistent way.  In this study, we use the methods developed in those works, with minor modifications.

The basic idea of KDE is to choose a global bandwidth $\beta$ %\eb{symbol already used for strain and also for latitude. keep h for strain and use other symbols for the other two} \js{fixed it with zeta for latitude and beta for global bandwidth} 
for a choice of kernel $K$, typically Gaussian, and given observations $X_i$ with corresponding weights $W_i$ estimate the density as
\begin{equation}
\label{eq:weighted_kde}
    \hat{f}(x) = \frac{1}{\sum_i W_i} \sum_{i=1}^{n} \frac{W_i}{\beta \lambda_i} K \left( \frac{x- X_i}{\beta \lambda_i} \right)\,. 
\end{equation}
Here, $\lambda_i$ is a local parameter multiplying the bandwidth for each observation~\cite{wang2011bandwidth}.  The expected error of the estimate has contributions from both its bias and variance: the choice of bandwidth is crucial to control these terms. Too small bandwidth, implying under-smoothing, causes the KDE to be too sensitive to small fluctuations in the data, inflating the variance, whereas over-smoothing occurs with large bandwidth, making the KDE unable to represent rapid variations in density and thus increasing its bias.  Furthermore, the optimal bandwidth at a given point in parameter space will depend on the density of observations in its neighborhood: smaller bandwidth is preferred in denser regions.

To address this issue, a per-point adaptive bandwidth is used: this is implemented by first evaluating a pilot KDE $\hat{f}_0(x)$, where we set $\lambda_i = 1$ for all events. 
Then choosing a sensitivity parameter $0 < \alpha \leq 1$ we determine the local adaptive bandwidth factor as 
\begin{equation}
 \lambda_i = \left( \frac{\hat{f}_0(X_i)}{g} \right)^{-\alpha}, \, \, \log g = n^{-1} \sum_{i=1}^{n} \log \hat{f}_0(X_i)\,,
\end{equation}
and finally evaluate the KDE via \eqref{eq:weighted_kde} using these $\lambda_i$ values. 

The global bandwidth $\beta$ and sensitivity parameter $\alpha$ remain to be chosen: we determine their optimal values by grid search, using the total log likelihood as a figure of merit. We evaluate the likelihood via \textit{K-fold cross-validation}, setting $K=5$ unless otherwise specified (see Ref.~\cite{Silverman1986} section 3.4.4 and~\cite{hastie01statisticallearning}).
The cross-validated (log) likelihood 
is
\begin{equation}
  \log \mathcal{L}_{\rm Kfold} = \sum_{k=1}^{K} \sum_{i \in \text{Fold}_k} \log \hat{f}_{{\rm Kfold},i}(X_i),
\end{equation}
where the data points are split into $K$ subsets of equal size: $\text{Fold}_k$ represents the $k$-th subset, and $\hat{f}_{{\rm Kfold},i}$ is the KDE constructed using the samples from the remaining $K-1$ subsets (excluding the points in $\text{Fold}_k$). 

We further apply an iterative reweighting method, introduced in \cite{Sadiq:2023zee}, in order to self-consistently treat measurement uncertainties in the detected events and avoid an over-dispersed estimate of the true distribution.  The motivation of this reweighting is as follows: posterior PE samples are produced using uniform or uninformative parameter priors, but we would obtain more accurate and precise measurements of astrophysically interesting parameters by instead using a prior close(r) to the true astrophysical distribution.  Conversely, with more accurate individual event parameters we would obtain a more accurate KDE for the population.  Our method addresses parameter errors by using the current estimate of KDE, obtained from some draw of samples from each event, to re-weight the samples of each event with probability proportional to the estimated density at each sample, in order to draw a new set of samples for the next bootstrap iteration. The new samples are then used to obtain a new KDE, and the process is repeated for many iterations, in a process similar in spirit to the expectation-maximization algorithm~\cite{10.2307/2984875}. We have shown that iterative reweighting can reduce the excess KDE dispersion due to measurement uncertainty~\cite{Sadiq:2023zee}. 

In our iterative process, the first sample reweighting uses a KDE based on the medians of 100 samples from each event. In subsequent iterations the KDE from the previous step is used for reweighting, producing a Markov chain of density estimates. 
We discard the first 100 iterations to exclude the initial transient behavior; after 100 additional iterations, at each step we then use the median of KDEs from a buffer containing the previous 100 iterations to derive sample weights for the next step's KDE, as explained in detail in~\cite{Sadiq:2023zee}.  We then collect 1000 bootstrap iterations using this median buffer reweighting, from which we calculate summary estimates via the median, 5th and 95th percentile of the KDEs. 

Our KDE derived from detected events is susceptible to selection bias, due to the detector's limitations in observing merging binaries over a finite range of masses and redshifts. It is crucial to account for selection effects to avoid biased estimates of the population properties. In the following section, we describe our treatment of selection effects via estimating the probability of detection.

\subsection{Selection effects and validation of PE}
\label{ss:pdet}
The probability of detection $p_{\rm det}$ for a given coalescing binary source is, in general, a function of all the parameters that we estimate (see Sec.~\ref{ss:pe_method}).  As we do not attempt detailed modelling of non-ideal data and the complexity of search pipelines, we use a relatively simple detectability criterion based on the SNR of simulated signals.  Specifically, we suppose that a search would detect an event if the matched filter SNR is above a given threshold $\rho_{\rm th}$ which we set to $8$.  Consistent with the treatment of PE to estimate measurement uncertainties for noiseless injections, we take the matched filter SNR $\rho$ to be a Gaussian with mean equal to the signal optimal SNR $\bar{\rho}$ and standard deviation of unity. (Strictly, for detection of a signal of unknown phase we should consider a non-central chi-squared distribution with 2 degrees of freedom~\cite{Maggiore2007_book,Essick:2023toz};
however, in practice the difference in $p_\mathrm{det}$ relative to the simple Gaussian is small.)  Thus, the probability of detection is the tail distribution function (complementary cumulative distribution function) of a unit Gaussian centered on $\bar{\rho}$, evaluated at $\rho_{\rm th}$. 

In a realistic situation, though, the true signal parameters are unknown: we do not have access to $p_{\rm det}$ for these parameters, but can only obtain an estimate of it based on the PE results. 
We thus begin with order($10^3$) posterior parameter samples for each event, obtained as described in Sec.~\ref{ss:pe_method}, and compute the optimal SNR for all samples.  
%We first compute the posterior SNR for ev intrinsic/extrinsic parameters and 
This enables us to validate the output of PE, as for a given event we expect a relatively narrow spread of posterior sample $\bar{\rho}$.  However, for a small number of events we find samples at very high redshift ($z >30$) and/or large mass ratio ($q >100$) having very low SNR ($<4$).  We interpret this as a failure of PE to converge to a posterior distribution reflecting the event's true parameters rather than the prior. 
To ensure data quality, we calculate the median and standard deviation of sample optimal SNR for each event. Events with a median SNR below 7 or a standard deviation above 2 were discarded, accounting for less than 4\% of the total dataset. Additionally, we filtered individual samples for the remaining events to remove those with an optimal SNR below 4.
 
To treat selection effects in our population analysis, 
we consider three distinct groups of source parameters: first, the KDE parameters $\{x\}$, i.e.\ those of most physical interest in reconstructing the population distribution, here comprising the source-frame total mass and redshift.  Quantifying the source distribution over $M$ and $z$ is crucial to understand the formation and evolution of MBH binaries and their role in galaxy formation.  Second, other intrinsic parameters $\{\theta\}$, which may contain some astrophysically relevant information, as they may depend on the binary formation and evolution mechanisms, and may influence $p_{\rm det}$: these include the mass ratio and component spins.  Lastly, extrinsic parameters $\{\psi\}$, which are expected to have uniform/isotropic or uninformative distributions from symmetry arguments: these comprise the source direction and orientation relative to the Solar System and the time of coalescence.     

Taking a uniform/isotropic distribution of sources over the time and angular parameters, $p(\psi)$, for given redshift and intrinsic parameters we can define the detection probability: 
\begin{equation} \label{pdet_xtheta}
    p_{\rm det}(x,\theta) = \int p(\rho>\rho_{\rm th}|\bar{\rho}(x, \theta, \psi)) p(\psi)\, d^n\psi\,,
\end{equation}
i.e.\ the probability of matched filter SNR to be above threshold given the optimal SNR for a binary with fixed $x, \theta$ but unknown $\psi$.  We compute this probability for the redshift, masses and spins of any given posterior sample via Monte Carlo integration over $\psi$, where for each draw of angular extrinsic parameters we also rescale the luminosity distance by a random weak lensing error given by Eq.~\eqref{eq:sigma_wl}.  

The astrophysical rate density of mergers over intrinsic parameters and redshift, $R(x,\theta)$ is connected to the rate density of detected events $R_{\rm det}(x,\theta)$ via $\mathbf{R_{\rm det}(x,\theta) = p_{\rm det}(x,\theta)R(x,\theta)}$.  The astrophysical density over $x$ requires marginalization over $\theta$ via $R(x) = \int R(x,\theta)\, d^n\theta$.  However, the density of \emph{detected} events over $x$, which we have direct access to via the KDE, is given by 
\begin{equation} \label{eq:Rdet}
    R_{\rm det}(x) = \int R(x,\theta) p_{\rm det}(x,\theta) \,d^n\theta\,.
\end{equation}
Here, the joint density $R(x,\theta)$ is \emph{a priori} unknown, and we are currently unable to reconstruct the full \emph{detected} distribution over $x,\theta$ due to its high dimensionality.  To proceed, either some simplifying assumptions are needed (in the absence of more sophisticated methods for multi-dimensional KDE), or a strategy which allows us to estimate $R(x)$ directly without considering the joint density.  

In previous works, it was assumed that the joint distribution factorizes as $R(x, \theta) = R(x) p_{\rm pop}(\theta)$: the ``additional'' intrinsic parameters $\theta$ were taken to follow a simple, fixed distribution~\cite{Sadiq:2021fin,Sadiq:2023zee} based on independent population studies (e.g.~\cite{LIGOScientific:2020kqk}).  Then Eq.~\eqref{eq:Rdet} factorizes, and we may define 
\begin{multline}
    R_{\rm det}(x) = R(x) \int p_{\rm pop}(\theta) p_{\rm det}(x,\theta) \,d^n\theta\, \\
    \equiv R(x) p_{\rm det}(x;p_{\rm pop})\,,
\end{multline}
enabling us to pass between the detected and astrophysical distributions via a function of $x$ alone.  (We would also obtain a simple conversion $p_{\rm det}(x)$ by neglecting the $\theta$-dependence of $p_{\rm det}(x,\theta)$, but this is unlikely to be a good approximation.) This approach is only as good as the accuracy of its assumptions, in particular that the joint distribution $R(x, \theta)$ is of product form. 

Thus here, we consider different strategies for reconstructing either the observed distribution, proportional to $R_{\rm det}(x)$, or the astrophysical distribution $R(x)$.  We expect to a first approximation that the distribution of posterior samples for detected events over $x$ and $\theta$ will mirror the detected distribution $R(x,\theta) p_{\rm det}(x,\theta)$.
Thus, if we apply weights $\propto 1/p_{\rm det}(x,\theta)$ to posterior samples, the resulting weighted distribution will mirror the astrophysical one.  We may make use of this in two ways.  

Our \emph{first method} is a reconstruction of the \emph{detected} distribution via a KDE of detected event samples, setting $W_i=1$ in~\eqref{eq:weighted_kde}.  When applying iterative reweighting, to improve the accuracy of event parameters we require the weights to be proportional to the astrophysical distribution $R(x)$~\cite{Sadiq:2023zee}. We approximate this weighting here by multiplying the previous iteration's detected event KDE by a factor $\sim\!1/p_\mathrm{det}$, evaluated at each sample. 

By contrast, our \emph{second method} uses a weighted KDE which incorporates selection effects as KDE weights $W_i \sim 1/p_{\mathrm{det},i}$, in order to obtain a estimate of the astrophysical distribution, at least over regions of $x$, $\theta$ where detections exist and $p_{\rm det}$ is not very small. 

A na\"ive application of such ``inverse $p_{\rm det}$'' weights can, though, run into problems of statistical stability and bias.  Consider moving on a trajectory in the KDE space $x$ such that $p_{\rm det}$ tends to zero, while the astrophysical distribution is approximately constant. Then to first order, the density of samples will also tend to zero, and our estimate of the astrophysical density will become undefined, in practice becoming excessively sensitive to small number fluctuations rather than approaching a constant.  Furthermore, we expect the posterior sample distribution to be somewhat broader (i.e.\ have wider support) than the actual distribution of detected events, due to PE measurement uncertainty.  Hence, a $1/p_{\rm det}$ weighted KDE may be liable to uncontrolled overestimates at the edge of the sample distribution. 

To obtain a well-defined KDE for the astrophysical distribution with variance under control, we must therefore accept some bias in regions where $p_{\rm det}$ approaches zero.  For instance, we can control the variance at the cost of allowing the weighted KDE to be an underestimate of the astrophysical distribution in regions with few or zero detected events: this can be achieved by removing samples with very small $p_{\rm det}$, or by capping the $1/p_{\rm det}$ weights at a maximum value. In our main analyses, we implement weighting by a factor $1/(\max(p_{\rm det}, 0.1))$: thus, we do not expect to reconstruct the astrophysical population well in regions with $p_{\rm det} \ll 0.1$.
 
\subsection{Application to simulated data}
\label{ss:applicationKDE}
For our analysis, we randomly choose 100 posterior samples from each event.  For each selected sample, we find the corresponding $p_\mathrm{det}(x, \theta)$ at the given mass, spins and redshift (distance) via a Monte Carlo over 1000 randomized extrinsic angular parameters and coalescence time values using the \textit{LISAbeta} package~\cite{marsat2024lisabeta, Marsat:2020rtl}, also applying a random weak lensing factor to the signal amplitude via Eq.~\eqref{eq:sigma_wl}, to compute the optimal SNR, with a matched filter SNR threshold $\rho_{\rm th}=8$. We then use the two methods of Section \ref{ss:pdet} to obtain a KDE over redshift $z$ and (log) source-frame total mass, $\log_{10}(M/M_\odot)$. 

Considering different strategies to improve the accuracy of our KDE and account for selection effects, as discussed in Sec.~\ref{ss:pdet}, we employ two distinct methods for population reconstruction. 
In our \emph{first method} we apply adaptive KDE, computed using \textsc{awkde}~\cite{awkdecodelink}, within the iterative framework described in Sec.~\ref{ss:iterkde} 
to reconstruct the detected event distribution: we then only consider selection effects in the reweighting of samples for each event during iterative estimation.  The global bandwidth $\beta$ and sensitivity parameter $\alpha$ are optimized at each iteration from a grid search, with $\beta$  ranging from 0.01 to 0.9 and $\alpha$ from 0.1 to 0.8. During iterative reweighting, the value of the previous KDE at each sample is multiplied by an inverse selection function, here taken as $1/\max(p_\mathrm{det}, 0.1)$. We also performed an analysis where the reweighting steps do not contain this selection factor, the results being presented in Appendix~\ref{app:awkde_without_pdetfactor}.

Our \emph{second method} aims to reconstruct the astrophysical distribution: we thus take into account selection effects using a weight $W_i$ derived from $p_\mathrm{det}$ for each given sample.  We implement the weighting via a standard KDE (without adaptive bandwidth) \cite{2020SciPy-NMeth}, using $1/\max(p_\mathrm{det}, 0.1)$ as weights for our main result. The only hyperparameter to be optimized is then the global bandwidth $\beta$, searching in a grid of values ranking from 0.01 to 0.9 as for the first method.  We compare the resulting weighted KDE with the true astrophysical distribution in the next section.  

To derive rate density estimates from our weighted KDEs, we count detected events and use sample weights. For regions with $p_\mathrm{det} \simeq 1$ we have a one-to-one correspondence between detected and astrophysical events. For $p_\mathrm{det} < 1$, we scale the astrophysical density relative to the detected event density by 1/$p_\mathrm{det}$ to account for reduced detection probability. Thus, each KDE is normalized by the sum of weights $\sum_i W_i$ to ensure proper scaling (note that the the normalization factor varies between iterations). 

\section{Results}
\label{sec:result-section}
Considering all injected events with SNR above the threshold of 8 in $4\,$yr of simulated LISA observations, we determine the optimal SNR for each posterior sample, using both the intrinsic and extrinsic sample parameters. As described in Section \ref{ss:pdet}, we exclude certain events with poor PE performance. From the remaining 326 events, only samples with an optimal SNR above 4 are retained. From this subset, we randomly select 100 samples per event and compute their detection probability $p_\mathrm{det}$ using a matched filter SNR threshold of 8. 
\begin{figure}[tbp]
    \vspace*{-0.15cm}
    \includegraphics[width=0.48\textwidth]{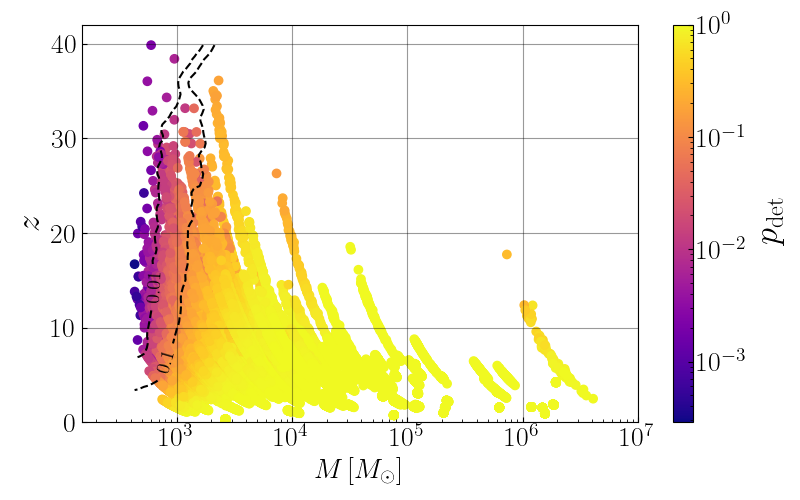}
    \caption{Detection probability $p_\mathrm{det}$ for 100 randomly chosen posterior samples from each detected event.
    Taking the intrinsic parameters and redshift of each posterior sample, we perform a Monte Carlo integration over the remaining extrinsic parameters and apply a matched filter SNR threshold of $8$ 
    to obtain $p_\mathrm{det}$. 
    We add contours (black dashed lines) to guide the eye based on these $p_{\rm det}$ values with nearest-neighbor interpolation and Gaussian smoothing.  
    %(with $\sigma=1$).
    }    \label{fig:Pdet_PESamplesLensedevent4years}
\end{figure}
Figure~\ref{fig:Pdet_PESamplesLensedevent4years} shows the total source-frame mass and redshift of the chosen samples from all events {passing our criteria}, colored according to $p_\mathrm{det}$.  {The range of redshifts covered by these samples is much wider than the true event distribution, reflecting the large PE uncertainties, and the range of source total masses correspondingly extends well below $10^3\,M_\odot$ in contrast to the true values shown in Fig.~\ref{fig:true-SNRdetectedLensedevent4years}.}

To test the accuracy of our methods, we 
compare to the properties of the underlying population, which is described in Sec.~\ref{ss:catalogdata}, focusing on the total mass and redshift of the mergers. 
For this comparison, we construct KDEs over (true) total mass and redshift values for the underlying population and its detectable sub-population, using \textsc{awkde}~\cite{awkdecodelink,Sadiq:2021fin}.  These reference KDEs represent the \emph{expected} distributions, and can be thought as averages over several realizations of actual 4\,yr merger catalogs.
In detail, the KDE global bandwidth $\beta$ and the $\alpha$ parameter are optimized with a grid search with $\beta$ values ranging from 0.01 to 1.0 and $\alpha$ values from 0 to 1. 
\begin{figure}[tbp]
    \centering
    \includegraphics[width=0.48\textwidth]{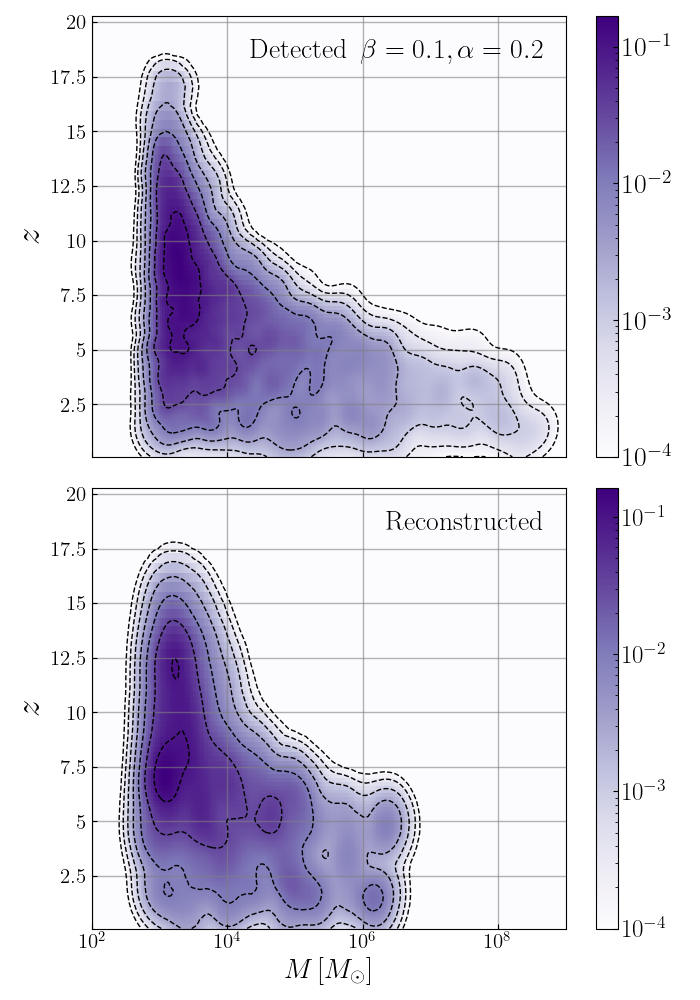}
    \caption{Top: KDE of the expected distribution over total mass and redshift, obtained using true parameter values sampled from the detectable subpopulation; $\beta$ and $\alpha$ are the optimized bandwidth and adaptive parameter respectively. Bottom: reconstructed distribution from the median of adaptive iterative
    KDEs from posterior samples of detected events over a 4 year observation time.  
    %\eb{I am not sure I understand this figure. the top panel is the astro pdf with no cut in snr (or so it seems) (..)}  \td{No, the top panel has the SNR cut, the top panel of fig 6 doesn't have it.}
    }
    \label{fig:firstKDEscomparison}
\end{figure}

\paragraph*{First method: Unweighted adaptive KDE for the detected rate density} Following Sec.~\ref{ss:applicationKDE}, we reconstruct the detected distribution over total mass and redshift from our posterior sample data, using adaptive but unweighted KDE. The results are shown in Fig.~\ref{fig:firstKDEscomparison}, where the median KDE (bottom) is compared with the KDE expected distribution of the detectable subpopulation (top): the two estimates closely resemble each other in most of the parameter space. However, the true expected distribution (top) exhibits structure at high masses $> 10^7\,M_\odot$ which is absent from the reconstruction, due to the lack of detected high-mass mergers in our mock data.  Conversely, at high redshift the range of mass covered by the reconstruction is slightly broader than the true detected distribution, likely due to the residual effect of PE uncertainties in source total mass. 
The iterative reweighting method reduces the impact of single-event measurement uncertainties, ensuring a more accurate and robust reconstruction, but may not be able to completely compensate for these effects.

\begin{figure}[tbp]
    \includegraphics[width=0.48\textwidth]{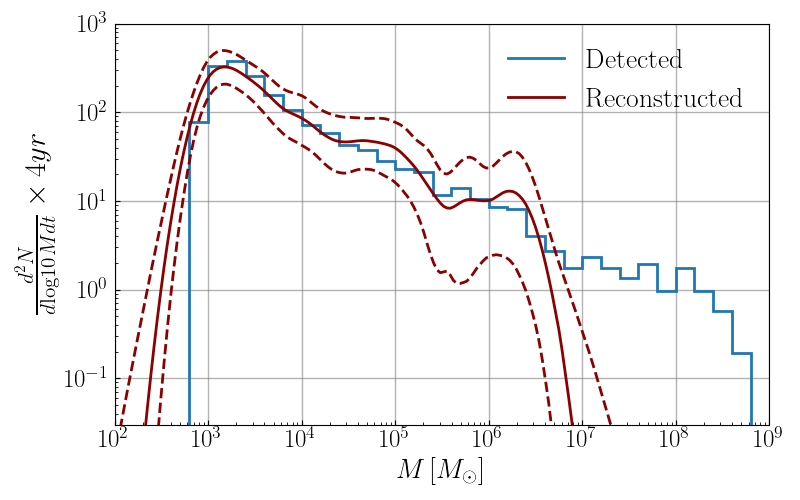}
     \includegraphics[width=0.48\textwidth]{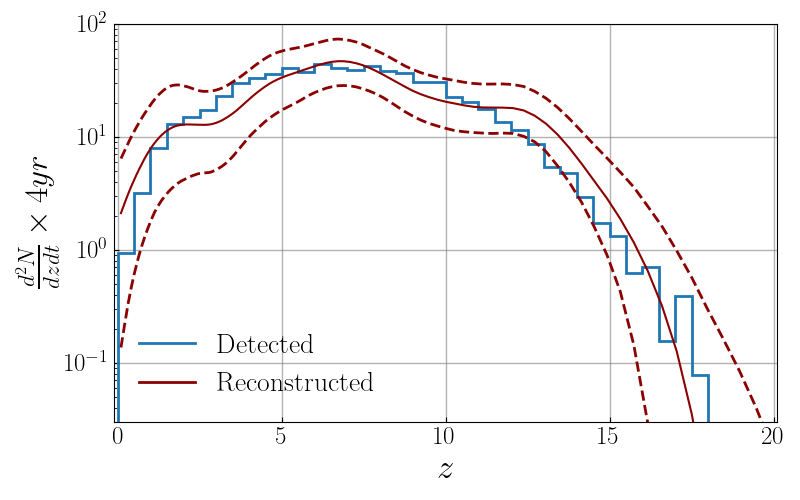}
     \caption{Rate densities over total source mass (top) and redshift (bottom) for 4 years of LISA observation.  We use adaptive unweighted %(i.e., without selection effects) 
     iterative KDE (dark red), plotting the median reconstruction (solid) and $90\%$ symmetric interval (dashed) compared with a histogram of true detected event values (light blue).}
    \label{fig:UnweightedIterativeOneDKDE}
\end{figure}

We convert the two-dimensional KDEs into one-dimensional KDEs by marginalizing over each variable in turn; one-dimensional detected event rate densities are then estimated by multiplying by the total number of observed events, 326. We compare the resulting reconstructed rates against the expected distribution of detectable events obtained from the astrophysical model in Fig.~\ref{fig:UnweightedIterativeOneDKDE}. These comparisons confirm the trends seen in the two-dimensional $M$--$z$ density, and indicate that the KDE reconstruction is generally accurate within the bootstrap uncertainties, except in parameter regions with zero/few detections; note that statistical uncertainties and fluctuations grow significantly in such regions, for instance around a total mass of  $\sim\!10^6\,M_\odot$.

As discussed in Section~\ref{ss:pdet}, we mitigate selection effects in the sample reweighting procedure by multiplying the detected event KDE by $1/\max(p_{\rm det}, 0.1)$, in order to approximate the underlying astrophysical distribution and thus improve the accuracy of individual event parameters. 
To assess the impact of this factor, we also conducted an analysis entirely without accounting for selection bias; the resulting rate estimates, presented in Appendix~\ref{app:awkde_without_pdetfactor}, confirm that this neglect leads to a significant downward bias in the redshift distribution. 

We also investigate the effect of using a simple $1/p_{\rm det}$ factor in reweighting (i.e.\ without capping $p_{\rm det}< 0.1$ values), with results shown in Appendix~\ref{app:awkde_with_pdetfactor_nocap}. While the overall trend of rate vs.\ total mass remains consistent with our main analysis, rates at large redshift values are now significantly overestimated, due to reweighting with factors $1/p_{\rm det} \gg 10$ in the presence of large redshift uncertainties, as discussed in Sec.~\ref{ss:pdet}.

\begin{figure}[tbp]
    \centering
    \includegraphics[width=0.48\textwidth]{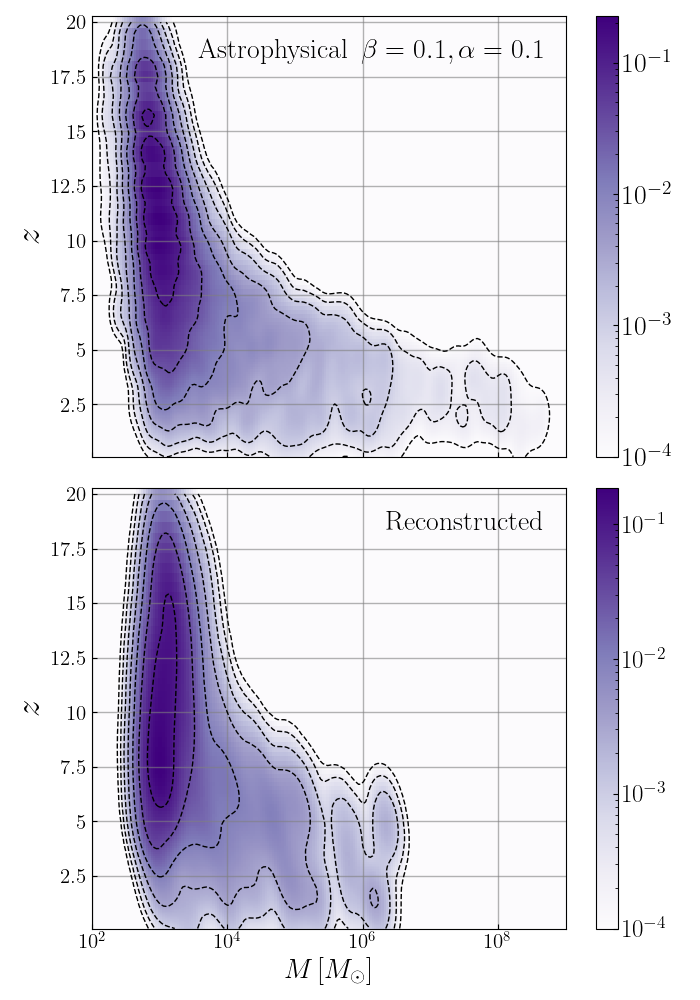}
    \caption{Top: KDE of the expected distribution of mergers over total source mass and redshift; $\beta$, $\alpha$ are the global bandwidth and local adaptive smoothing parameter respectively.
    Bottom: reconstructed distribution from the median of adaptive weighted KDEs from posterior samples of detected events over a 4 year observation time.
    }
    \label{fig:firstweightedKDEscomparison}
\end{figure}

\paragraph*{Second method: Weighted KDE for the astrophysical distribution} 
Here, we follow Sec.~\ref{ss:applicationKDE},  incorporating selection effects via weights $W_i = 1/\max(p_{\rm det, i}, 0.1)$ applied to the posterior samples in order to estimate the underlying astrophysical distribution. For comparison, an adaptive KDE of true mass and redshift value from the full astrophysical population is shown in Fig.~\ref{fig:firstweightedKDEscomparison} (top).

The reconstructed distribution from iterative weighted KDE is shown in the bottom plot of Figure~\ref{fig:firstweightedKDEscomparison}, showing similar behavior {over most of} the $M$–$z$ plane. The KDE sample weights
effectively account for selection effects at low mass and high redshift, demonstrating their crucial role in population reconstruction.     
As with the first method, we see higher statistical fluctuations and ultimately a lack of support in the KDE reconstruction towards the high mass limit $M \sim 10^7\,M_\odot$, due to the scarcity of detected events in this range.  The high-redshift astrophysical density at $M< 10^3\,M_\odot$ is also not perfectly represented by the weighted KDE, as the detection probability drops well below $0.1$ here.

\begin{figure}[tbp]
    \includegraphics[width=0.48\textwidth]{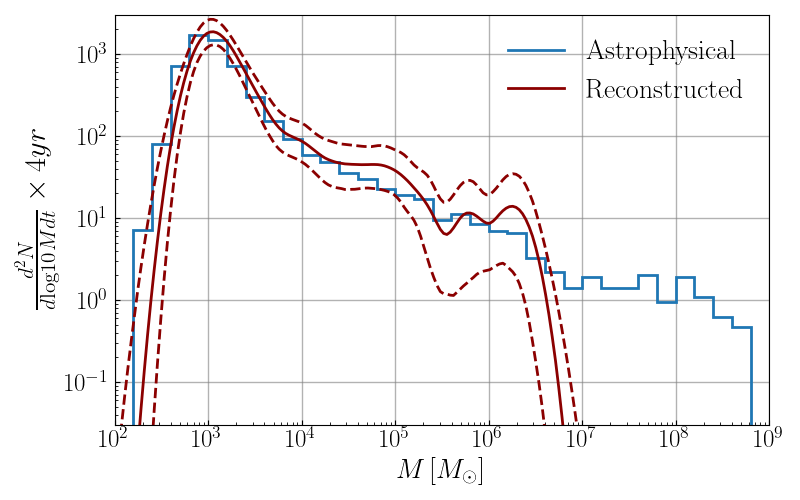}
     \includegraphics[width=0.48\textwidth]{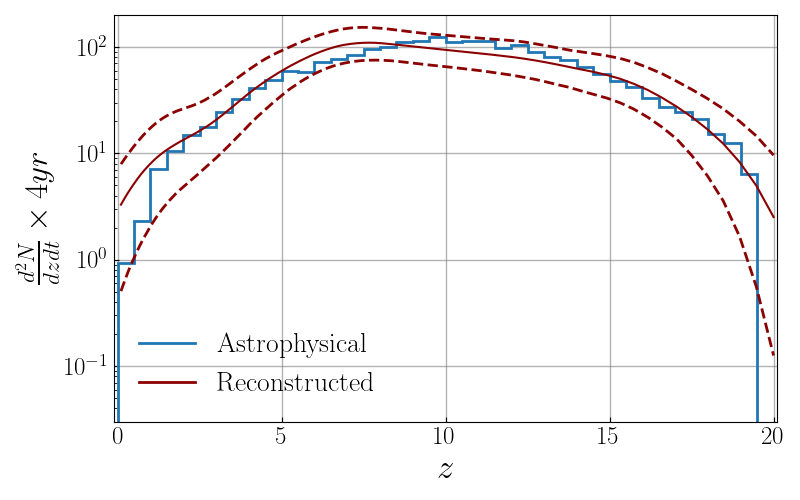}
 \caption{Rate densities over total source mass (top) and redshift (bottom) for 4 years of LISA observation.  We use weighted %(i.e., without selection effects) 
 iterative KDE (dark red), plotting the median reconstruction (solid) and $90\%$ symmetric interval (dashed) compared with a histogram of true values for the whole astrophysical population (light blue).
}
    \label{fig:weightedIterativeOneDKDE}
\end{figure}

From the two-dimensional reconstructed KDE, we compute one-dimensional KDEs as before. We then 
use sample weights to calculate event-based rates as explained in Sec.~\ref{ss:applicationKDE}. The reconstructed one-dimensional rate estimates are compared with the histogram of astrophysical rates, as shown in Fig.~\ref{fig:weightedIterativeOneDKDE}, showing trends consistent with the two-dimensional comparison, including higher uncertainties around $M \sim 10^6\,M_\odot$. 

To further explore the impact of selection effects %$p_{\rm det}$ 
in our second method, we performed an alternative analysis with weights simply given by $W_i = 1/p_{\rm det, i}$, without truncation of small $p_{\rm det}$ values. The results as presented in  Appendix~\ref{app:weightedKDEwithsmallpdet} show some noticeable deviation from the astrophysical distribution: we trace this to a preference for much larger global bandwidths $\beta$, resulting from the fact that the weighted KDE is strongly influenced by a small number of samples with $W_i \gg 1$ close to the low-mass edge of the distribution.  Effectively, the bulk of the population must then be fitted by contributions from such ``edge-case'' events, but this cannot be achieved without significant bias elsewhere.  This problem suggests that the dynamic range of KDE weights must be limited to avoid domination of the estimate by very high values. 

\begin{figure}[tbp]
 \includegraphics[width=0.48\textwidth]{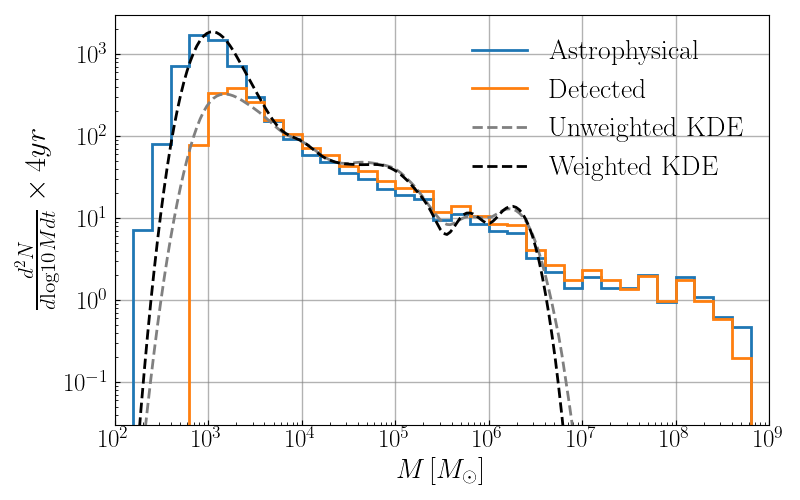}
     \includegraphics[width=0.48\textwidth]{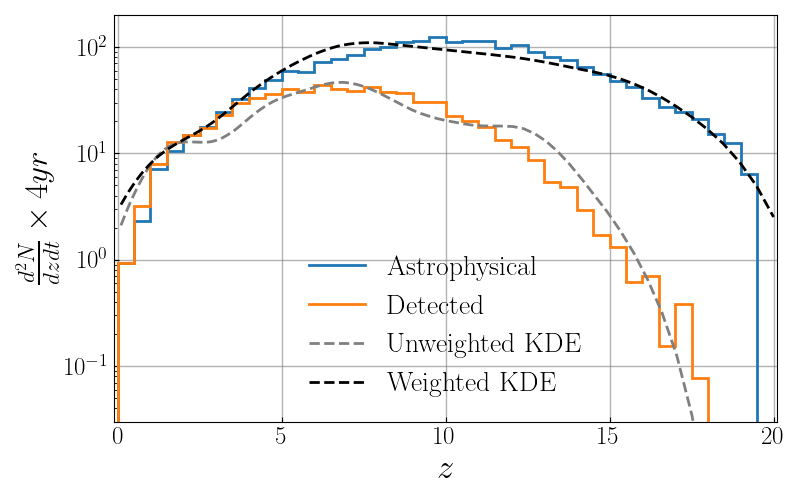}
 \caption{Rate densities over total source mass (top) and redshift (bottom) for 4 years of LISA observation time.  The median weighted iterative KDE (dashed black) is compared with the expected astrophysical distribution (light blue), whereas the unweighted iterative KDE (dashed gray) is compared with the expected distribution of detectable events (orange). 
}
    \label{fig:finalcompare}
\end{figure}

Our results for both the adaptive unweighted KDE and weighted KDE methods, using the choice of (re)weighting factor $1/\max(p_{\rm det}, 0.1)$ to account for selection effects, are summarized in Fig.~\ref{fig:finalcompare}, where for simplicity we show only the median KDE rate estimate.   
 
\section{Conclusion}
\label{sec:conclusion-section}
%\eb{
Understanding the properties of the astrophysical population of binary compact objects throughout the universe is a fundamental aim of GW observatories. However, this task is complicated by the inherent limitations in the strain sensitivity, frequency range, and observation duration of both present and future detectors. As a result, we are typically able to observe only a fraction of the total population, with limited event rates and significant uncertainties in the measured parameters of each detected event. In this paper, we concentrate on (simulated) detections of massive black hole binaries by LISA. If these black holes originate from population III star remnants, or ``light seeds'', a large portion of the low-mass, high-redshift binary population could remain undetected. As a result, selection biases must be carefully considered and corrected when reconstructing the underlying population properties from the observed data. Similar problems apply to existing and future terrestrial detectors (see e.g.~\cite{Pieroni:2022bbh}).

In this paper, we have proposed and tested a non-parametric kernel density estimation (KDE) method with iterative reweighting 
%based on expectation maximization algorithm, 
aimed at addressing these challenges. Our approach tackles two key aspects: selection effects that limit the detectability of certain signals, and the inherent statistical uncertainties in the parameters of observed events. 
 
We employed two methods in our analysis.
In the first, we use adaptive KDE without correcting for selection effects directly. Instead, these effects are applied in 
the iterative reweighting process 
in order to improve the accuracy of event parameters. The resulting reconstructed rate estimates are then compared to the expected distribution of detected events drawn from the underlying astrophysical model: the detected distribution is recovered across most of parameter space, except in regions where no events are detected. 
Significant biases would arise either from ignoring selection effects, or by naively attempting to correct such effects without considering the stability of estimates in regions with very small detection probability $p_{\rm det}$.

In our second method, we use a weighted KDE, with weights designed to compensate for selection effects, and compare our reconstructed rate estimate with the astrophysical model distribution. Our results show that this approach can effectively reconstruct the astrophysical distribution when regularization is applied to the detection probability $p_{\rm det}$. However, without proper capping on small values of $p_{\rm det} \ll 0.1$, the weighted KDE becomes severely biased as the estimate is dominated by a minority of samples with very high weights. 
The regularization used here, while apparently effective, is somewhat ad-hoc and deserves further investigation: for instance, determining how the capping of weights should scale with the number of events and the size of their parameter uncertainties.

Our iterative KDE method has the advantage of being fast and flexible, and it can account for PE uncertainties. It can also complement and improve parametric models by providing non-parametric estimates that can validate or refine model assumptions. 
So far, we have deployed the method only over a one- or two-dimensional parameter space, with the restriction that the kernel is rotationally symmetric in the two-dimensional space (for standardized data).  This approach is suitable only for cases where the relative measurement uncertainties and scales of structure in the population density are comparable between parameters; the immediate further development required to generalize the KDE kernel would be allowing independent bandwidths for different parameters, with a suitable optimization method.  In general, density estimation over a higher-dimensional parameter space will also require a larger data set for a stable result (the ``curse of dimensionality'').
Ideally, it would be preferable to reconstruct the full distribution over the principal binary intrinsic parameters affecting detectability (masses and orbit-aligned spins) and redshift; 
%however at present we are limited by technical aspects of the KDE, in particular the choice of kernel for high-dimensional parameter spaces, 
we leave this as a goal for future work. 

\section*{Acknowledgements}
E.B.\ and J.S.\ acknowledge support from the European Union’s H2020 ERC Consolidator Grant ``GRavity from Astrophysical to Microscopic Scales'' (Grant No. GRAMS-815673), the PRIN 2022 grant ``GUVIRP - Gravity tests in the UltraViolet and InfraRed with Pulsar timing'', and the EU Horizon 2020 Research and Innovation Programme under the Marie Sklodowska-Curie Grant Agreement No. 101007855. 
K.D.\ acknowledges IISER Thiruvananthapuram for providing high performance computing resources at HPC Padmanabha.
T.D.\ is supported by research grant PID2020-118635GB-I00 from the Spanish Ministerio de Ciencia e Innovaci{\'o}n and also received financial support from Xunta de Galicia (CIGUS Network of research centers) and the European Union. 
\bibliography{reference}% Produces the bibliography via BibTeX.

\clearpage
\newpage

\appendix
\section{Results from adaptive KDE reweighting without selection factor}
\label{app:awkde_without_pdetfactor}

\begin{figure}[bhp] 
    \includegraphics[width=0.48\textwidth]{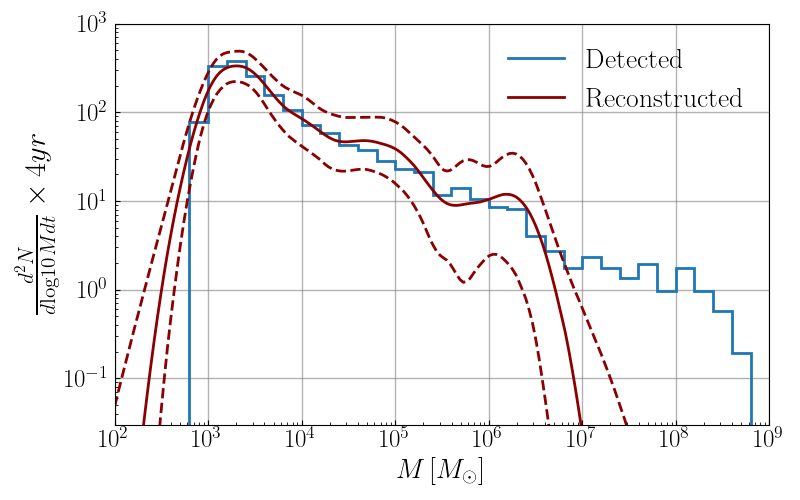}
     \includegraphics[width=0.48\textwidth]{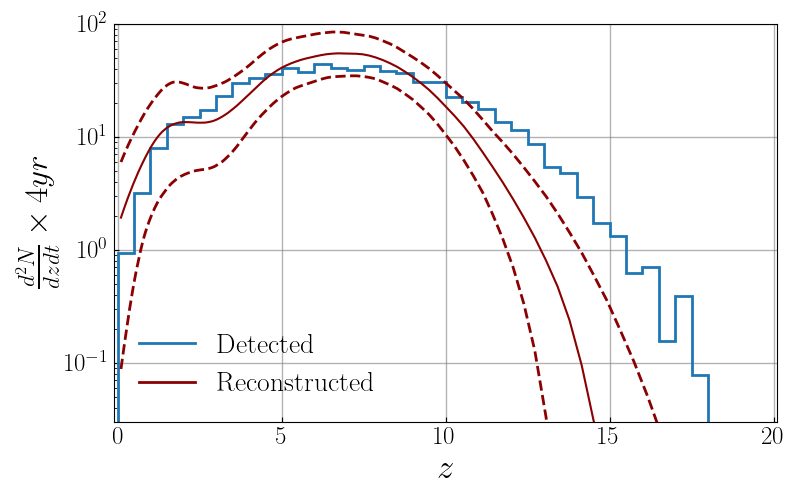}
    \caption{Rate densities over total source mass (top) and redshift (bottom) for 4 years of LISA observation time. 
    The reconstruction (solid dark red: median, dashed dark red: 90\% symmetric interval) is performed using adaptive but unweighted KDE, neglecting all selection effects.
    The light blue histogram represents the distribution of the population's detectable events.  
    }
    \label{fig:awkde_withoutpdetfactor}
\end{figure}

In addition to our main results, we perform an iterative adaptive KDE reconstruction without including selection effects in the reweighting, thus completely neglecting selection bias. 
Our results are compared to the expected distribution of detectable events in Fig.~\ref{fig:awkde_withoutpdetfactor}. 

While the marginal mass distribution is scarcely affected by neglecting selection effects, the redshift distribution is significantly biased towards lower values relative to the true distribution of detectable events.  Given the large redshift uncertainties, the method underestimates the redshift of events above $z\simeq 10$.  The bulk of detected events are at lower redshifts, and here this trend is (wrongly) attributed to a property of the underlying population rather than being due to selection. Hence, selection bias is crucial even to estimation of the detected distribution.

\section{Results from adaptive KDE using simple 1/$p_\mathrm{det}$ reweighting
%with without capping $p_\mathrm{det} < 0.1$}
}
\label{app:awkde_with_pdetfactor_nocap}

\begin{figure}[hbtp]
    \includegraphics[width=0.48\textwidth]{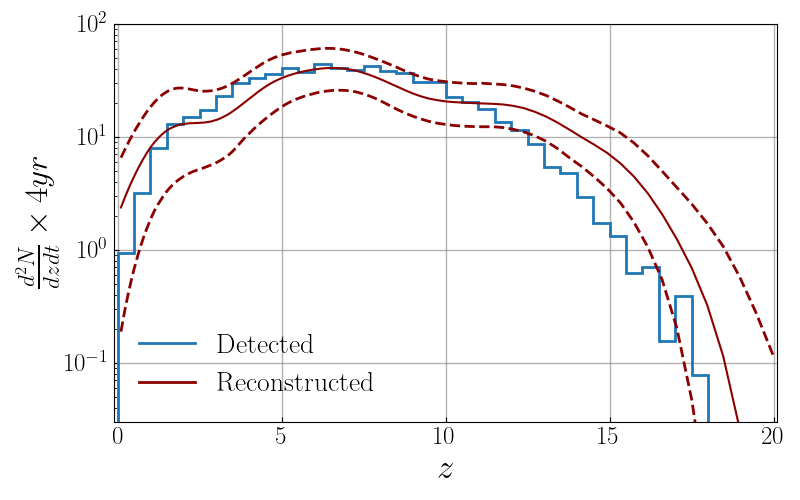}
     \includegraphics[width=0.48\textwidth]{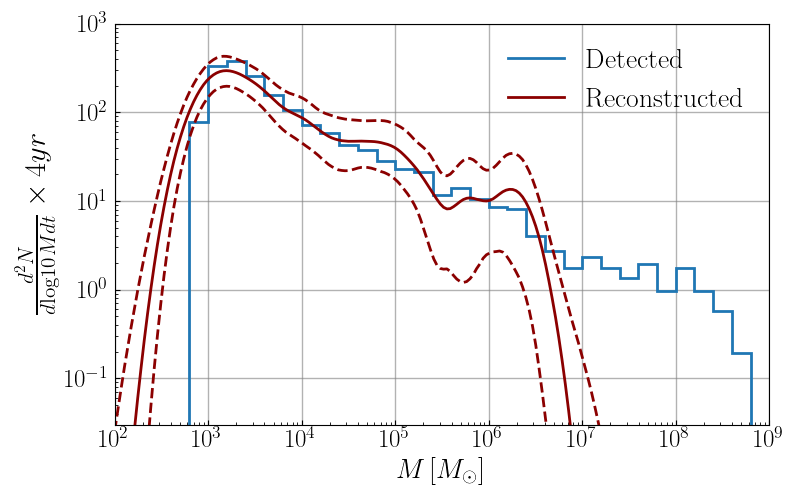}
     \caption{Rate densities over total source mass (top) and redshift (bottom) for 4 years of LISA observation time.  Solid (dashed) dark red lines show the median ($90\%$ symmetric interval) iterative KDE reconstruction from posterior samples: here, the iterative reweighting provides an estimated astrophysical population given by the detected event KDE multiplied by $1/p_{\rm det}$, without truncating very small values $p_{\rm det} < 0.1$.}
    \label{fig:awkde_withpdetfactor_nocap}
\end{figure}

Here, we perform an iterative adaptive KDE reconstruction including selection effects in the sample reweighting, but without limits on small $p_\mathrm{det}$ values. In more detail, at each 
step we draw new samples with weights given by the previous detected KDE multiplied by $1/p_\mathrm{det}$: we compare our results to the distribution of detectable events as shown in Fig.~\ref{fig:awkde_withpdetfactor_nocap}. 

We see here that including samples with very small values of $p_\mathrm{det} \ll 0.1$ and simply weighting by $1/p_\mathrm{det}$ significantly overestimates rates at high redshift values.  

\newpage 
\section{Results from weighted KDE using simple 1/$p_\mathrm{det}$ weights}
\label{app:weightedKDEwithsmallpdet}

In addition to our main weighted KDE results, we here perform a weighted KDE reconstruction including selection effects via weights $W_i$, defined without capping smaller values of $p_{{\rm det},i} < 0.1$, i.e.\ $W_i = 1/p_{\rm det , i}$. 

\begin{figure}[hbtp]
\includegraphics[width=0.48\textwidth]{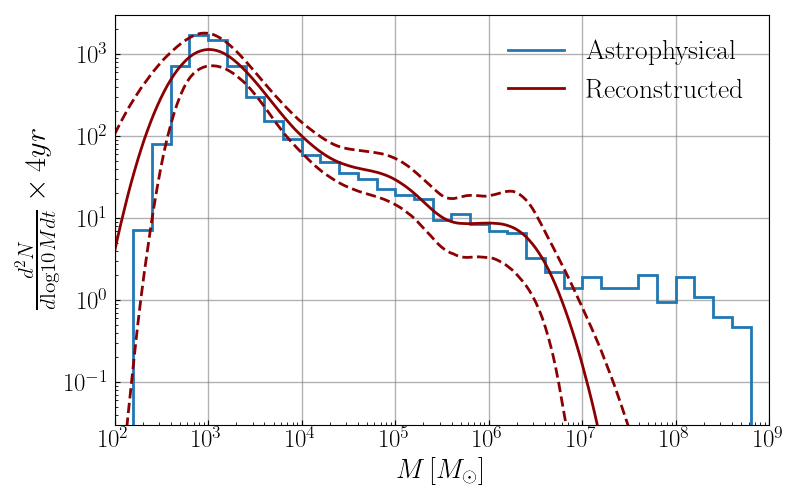}
\includegraphics[width=0.48\textwidth]{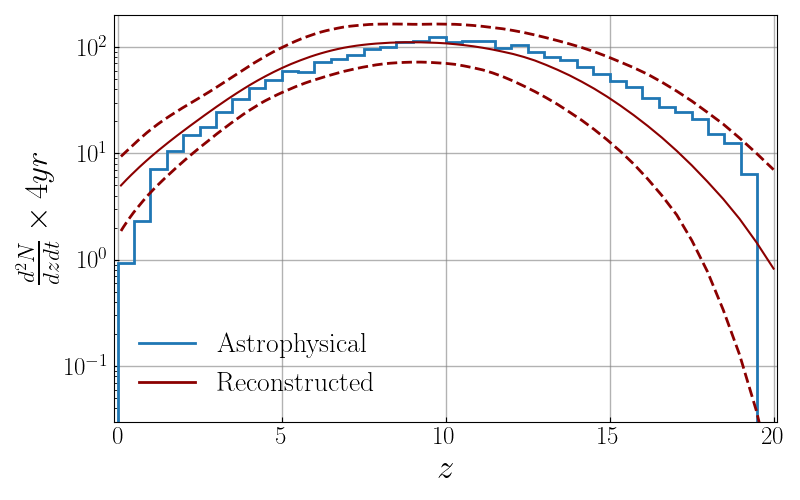} 
\caption{Rate densities over total source mass (top) and redshift (bottom) for 4 years of LISA observation time.    
The reconstruction (solid dark red, with $90\%$ symmetric interval in dashed dark red) is performed using weighted iterative KDE correcting for selection effects. The light blue histogram represents the distribution of the whole astrophysical event population.
}
    \label{fig:weighted_smallpdet}
\end{figure}
Comparing with the true astrophysical distribution in Fig.~\ref{fig:weighted_smallpdet}, we find some biases, particularly at low mass and both low and high redshifts.  The method now prefers much higher KDE bandwidths, an effect which can be attributed to the estimate being dominated by (the kernels of) a small number of samples with very high $W_i$ around the low-mass edge seen in Fig.~\ref{fig:Pdet_PESamplesLensedevent4years}: these can only fit the rest of the population if their bandwidth is inflated.  Hence, the KDE is forced to take a broad Gaussian form with few additional features.  This form can, coincidentally, reproduce the true $z$ distribution relatively well, as that happens to also be close to Gaussian, but would be unable to reconstruct any more complicated astrophysical population.

\end{document}